\documentclass[fleqn,twoside]{article}
\usepackage{espcrc2}

\usepackage{graphicx}
\usepackage[figuresright]{rotating}

\newcommand{\AmS}{{\protect\the\textfont2
  A\kern-.1667em\lower.5ex\hbox{M}\kern-.125emS}}

\title{\vspace*{-1.5cm} $Au$+$Au$ Collisions at RHIC and implications for ultra-relativistic
astrophysical $A$+$A$ collisions}

\author{Jens S\"oren Lange
\address[addr]{
Johann Wolfgang Goethe-Universit\"at Frankfurt am Main, Institut f\"ur Kernphysik\\
\quad\\
Presented at the XIII International Symposium on Very High Energy Cosmic Ray Interactions\\
Pylos, Greece, 09/6-12/2004
}
\thanks{Current Address: GSI, Planckstra\ss{}e 1, 64291 Darmstadt, Germany, Email $<$soeren@bnl.gov$>$}
}

\begin{document}

\begin{abstract}
\noindent
Results from ultra-relativistic $Au$+$Au$ collisions at RHIC are reviewed.
Emphasis is put upon {\it (a)}~measured properties of a $Au$+$Au$ collision,
which might be used as input to cosmic air shower Monte-Carlo event generators,
{\it (b)}~production of anti-matter, and 
{\it (c)}~forward physics.
\vspace{1pc}
\end{abstract}

\maketitle

\linepenalty=10
\widowpenalty=10
\clubpenalty=10


\section{RHIC}

\begin{figure}[hhh]
\label{f:rhic}
\centerline{\includegraphics[width=7.5cm]{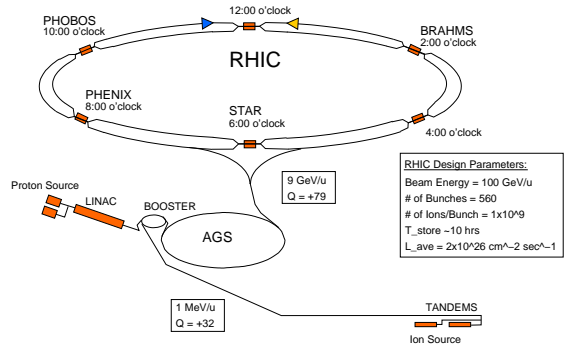}}
\vspace*{-0.3cm}
\caption{The RHIC accelerator complex.}
\end{figure}

\noindent
The Relativistic Heavy Ion Collider (RHIC)
at Brookhaven National Laboratory, New York, USA,
has a circumference of 3833~m and uses
1740 superconducting magnets \cite{rhic_overview}.
Fig.~\ref{f:rhic} shows the RHIC accelerator complex.
The main objective of RHIC is the investigation 
of novel QCD phenomena at high density and high temperature
in ultra-relativistic $Au$+$Au$ collisions.
The highest center-of-mass energy\footnote{Throughout this paper,
$\sqrt{s}$ denotes the center-of-mass energy in the nucleon-nucleon 
system $\sqrt{s_{NN}}$.} 
$\sqrt{s}$=200~GeV is a factor $\simeq$10 higher than past
fixed target nucleus-nucleus collision experiments (e.g.\ CERN SPS $E_{beam}$=160~GeV,
equivalent to $\sqrt{s}$$\simeq$17.3~GeV).
With a cross section of $\sigma_{Au+Au}$=7.04~barn (see Tab.~2),
the design luminosity of $\cal{L}$=2$\cdot$10$^{26}$~cm$^{-2}$s$^{-1}$
means an interaction rate of $\simeq$1.4~kHz.
Four experiments are located at RHIC interaction points, 
i.e.\ the STAR, PHENIX, PHOBOS and BRAHMS experiments.

\noindent
The STAR experiment consists of

\setlength{\parskip}{-0.0ex}%

\noindent $\bullet$ 
a large scale midrapidity ($|$$\eta$$|$$\leq$1.6) Time Projection Chamber 
(TPC, $R$=2~m, $L$=4~m)  with $\simeq$48,000,000 pixels
(momentum resolution $\Delta$$p_T$/$p_T$=3\% at $p_T$=1~GeV/c
in a solenodial magnetic field of $B$=0.5~T) \cite{star}

\noindent $\bullet$ 
2400 lead-scintillator electromagnetic calorimeter modules ($|$$\eta$$|$$\leq$1.0), and 

\noindent $\bullet$ 
a 3-layer silicon drift detector with $\simeq$13,000,000 pixels ($|$$\eta$$|$$\leq$1.0) 

\noindent
The PHENIX experiment \cite{phenix} consists of

\noindent $\bullet$ 
a midrapidity ($|$$\eta$$|$=$\pm$0.35) electron and hadron detection system,
using a ring imaging Cerenkov detector, a time-of-flight system, drift chambers
(mass resolution $\Delta$$m$/$m$=0.4\% at $m$=1~GeV)
and 24,768 lead-scintillator electromagnetic calorimeter modules, and

\noindent $\bullet$ 
forward/backward (1.1$\leq$$|$$\eta$$|$$\leq$2.2) muon chambers.

\noindent
The PHOBOS experiment \cite{phobos} uses 135,168 silicon strip and silicon pad readout channels in a coverage of
$|$$\eta$$|$=$\pm$5.4, $\Delta$$\varphi$=360$^o$, for unidentified charged particles, and
$|$$\eta$$|$=$\pm$2.0, $\Delta$$\varphi$=180$^o$$\pm$20$^o$ for identified charged particles.

\noindent
The BRAHMS experiment \cite{brahms} is a 
spectrometer with a small opening angle of $\Delta$$\Omega$=7.3~msr,
however at an adjustable rapidity position, thus providing particle identification 
from $|$$\eta$$|$=0 to $|$$\eta$$|$=4.

\noindent
As an example for all experiments, Tab.~1 shows the number of recorded
events at the STAR experiment since start of RHIC operation.

\quad\\
\vspace*{0.1cm}
\begin{tabular}{@{}llll}
\multicolumn{4}{l}{\hspace*{-0.2cm}Table 1}\\
\multicolumn{4}{l}{\hspace*{-0.2cm}Recorded STAR data set.}\\
\hline
 & $\sqrt{s}$ & $\sharp$ of Events & Year \\
\hline
$Au$+$Au$ & 130.0~GeV & 0.7~Mill. & 2000 \\
$Au$+$Au$ & 200.0~GeV & 3.2~Mill. & 2001 \\
$Au$+$Au$ & 19.6~GeV & $\simeq$20.000 & 2001 \\
$d$+$Au$ & 200.0~GeV & 35~Mill. & 2003 \\
$Au$+$Au$ & 200.0~GeV & 77.9~Mill. & 2004 \\
$Au$+$Au$ & 62.4~GeV & 13.3~Mill. & 2004 \\
$p$+$p$ & 200~GeV & 19.8~Mill. & 2004 \\
\hline
\end{tabular}
\vspace*{0.4cm}

\noindent
$Au$+$Au$ collisions create a system of {\it hot} nuclear matter,
$d$+$Au$ collisions a system of {\it cold} nuclear matter.
The $p$+$p$ data set serves as reference for $Au$+$Au$, but at the same
time is also used for the RHIC spin program, as protons in RHIC 
can be polarized vertical or longitudinal.

\noindent
As an illustration of the amount of data,
the raw data set collected at STAR experiment in 2004
compares to $\simeq$60 times the combined raw data set of 
the $B$ meson factories BELLE and BABAR.

\begin{figure}[hhh]
\label{f:event}
\hspace*{0.2cm}
\begin{center}
\framebox{\includegraphics[width=6.5cm]{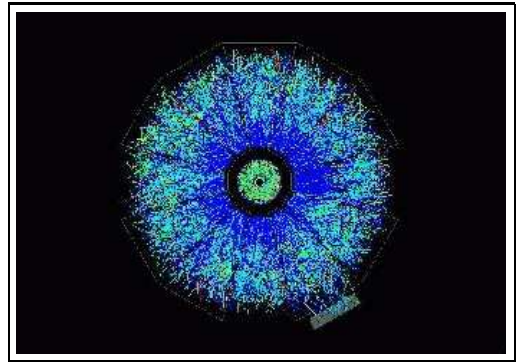}}
\end{center}
\vspace*{-0.9cm}
\caption{RHIC $Au$+$Au$ collision at $\sqrt{s}$=200~GeV, recorded at the STAR experiment.}
\end{figure}

\noindent
STAR also utilizes a level-3 trigger system 
\cite{star_l3}, performing full event reconstruction
of a central $Au$+$Au$ collision within $t$$\leq$100~ms.
Fig.~\ref{f:event} shows a central $Au$+$Au$ collision at $\sqrt{s}$=200~GeV, 
recorded by the level-3 trigger system at STAR, 
and consisting of $\simeq$6,500 charged
particle tracks and $\simeq$130,000 TPC clusters.
The level-3 trigger was used for realtime rare probe detection,
such as events with anti-Helium 
(described in detail in Ch.~\ref{ch:antihe})

\section{The RHIC $Au$+$Au$ Collision}

\label{ch:1coll}

\noindent
There are four basic stages of an $Au$+$Au$ collision,
which are also schematically depicted in Fig.~3.

\noindent $\bullet$ 
$t$=0~fm/c. The nuclei are Lorentz contracted in the laboratory frame (``pancake shape'').
Hard parton scattering occurs ($q$$q$, $q$$g$, $g$$g$ scatterings).

\noindent $\bullet$ 
$t$$\simeq$1~fm/c. A hot cylinder is formed. The temperature in the cylinder exceeds
10$^{12}$~K. 

\noindent $\bullet$ 
$t$$\simeq$4~fm/c. Soft partons, thermally produced, emerge from the center of the collision.
Beam remnant particles are peaked in the forward and backward directions.

\noindent $\bullet$ 
$t$$\simeq$10~fm/c. Hadronic freeze-out occurs. All partons confine into particles. 

\quad\\
\begin{tabular}{@{}llll}
\multicolumn{4}{l}{\hspace*{-0.2cm}Table 2}\\
\multicolumn{4}{l}{\hspace*{-0.2cm}Measured parameters of a RHIC $Au$+$Au$ collision.}\\ 
\multicolumn{4}{l}{\hspace*{-0.2cm}For details see Ch.~\ref{ch:1coll}}\\
\hline
 & & $\sqrt{s}$ & Ref.\ \\
\hline
$\sigma_{Au+Au}$ & (7.05$\pm$0.05)~barn & 130 & \cite{kharzeev2}\\
$N_{part}$ & 352$\pm$3 & 200 & \cite{star_Et} \\
$N_{bin}$ & 1051$\pm$72 & 200 & \cite{star_Et} \\
$R_{out}$  & (5.39$\pm$0.18$\pm$0.28)~fm & 130 & \cite{star_hbt} \\
$R_{long}$ & (5.99$\pm$0.19$\pm$0.36)~fm & 130 & \cite{star_hbt} \\
$R_{side}$ & (5.48$\pm$0.15$\pm$0.30)~fm & 130 & \cite{star_hbt} \\
$E_T$ & (620$\pm$33)~GeV & 200 & \cite{star_Et} \\
$E_T^{em}$ & (216$\pm$14)~GeV & 200 & \cite{star_Et} \\
$\beta$ & 0.56$c$ & 130 & \cite{star_multistrange} \\
$T_{chem}$ & (181$\pm$8)~MeV & 130 & \cite{star_multistrange} \\
$T_{kin}$ & (89$\pm$10)~MeV & 200 & \cite{star_lee2} \\
$\mu_b$/T & 0.18$\pm$0.03 & 130 & \cite{star_mub} \\
$\overline{p}$/$p$ ratio & 0.71$\pm$0.002$\pm$0.05 & 130 & \cite{star_antip} \\
$\varepsilon_{Bjorken}$ & (4.2$\pm$0.3)~GeV/fm$^3$ & 130 & \cite{star_lee1} \\
\hline
\end{tabular}\\[2pt]
\vspace*{0.4cm}

\noindent
Tab.~2 lists the measured parameters of a RHIC collision.
$\sigma_{Au+Au}$ denotes the total cross section.
$N_{part}$ denotes the number of participant nucleons.
$N_{bin}$ denotes the number of binary nucleon-nucleon collisions.
The size of the system can be concluded from the radii
$R_{out}$, $R_{long}$ and $R_{side}$, which are measured
by quantum-mechanical Hanbury-Brown-Twiss (HBT) interferometry
of $\pi^{\pm}$$\pi^{\pm}$ pairs, close in phase space.
$R_{out}$ denotes the radius perpendicular to the beam axis,
$R_{long}$ parallel to the beam axis.
$R_{side}$ denotes a radius perpendicular to $R_{out}$, 
but not necessarily measured from the origin (0,0,0).
$R_{out}$ can be interpreted as the size of the
particle wave front, while freezing out. 
In particular, the ratio $R_{out}$/$R_{side}$$\simeq$1
indicates that RHIC freeze-out has a surprisingly short duration,
an evidence which is supported by the measured high
expansion velocity, given as the
fraction of the speed of light $\beta$=0.56.

\noindent
$E_T$ denotes the total transverse energy,
defined as the sum of the electromagnetic transverse energy
$E_T^{em}$ (i.e.\ leptons and photons)
and the hadronic transverse energy 
$E_T^{had}$.
The ratio $E_T^{em}$/$E_T$ was determined to be 0.348$\pm$0.019 \cite{star_Et}.
So far, there is no indication for Centauro type events
with an anomalous ratio
$E_T^{em}$/$E_T^{had}$$\simeq$1/6 \cite{centauro}.

\noindent
The $\overline{p}$/$p$ ratio visualizes how many baryons from the beam projectiles
remain in the collision zone. A complete baryon-free zone would lead
to a $\overline{p}$/$p$ ratio $\simeq$1, i.e.\ all baryons and anti-baryons are 
created by pair production. The measured
$\overline{p}$/$p$ ratio of 0.71
indicates that a significant excess of baryons 
over anti-baryons is still present at RHIC. 
As a comparison, the measured ratio is much higher than 
0.00025$\pm$10\% at the AGS ($\sqrt{s}$$\simeq$4.9~GeV) and
0.07$\pm$10\% at the SPS ($\sqrt{s}$$\simeq$17.3~GeV),
i.e.\ RHIC collisions represent for the first time
an opportunity to advance
into the almost baryon free regime.

\begin{figure}[htb]
\label{f:time}
\vspace*{1ex}
\begin{center}
\framebox{\includegraphics[width=3.1cm]{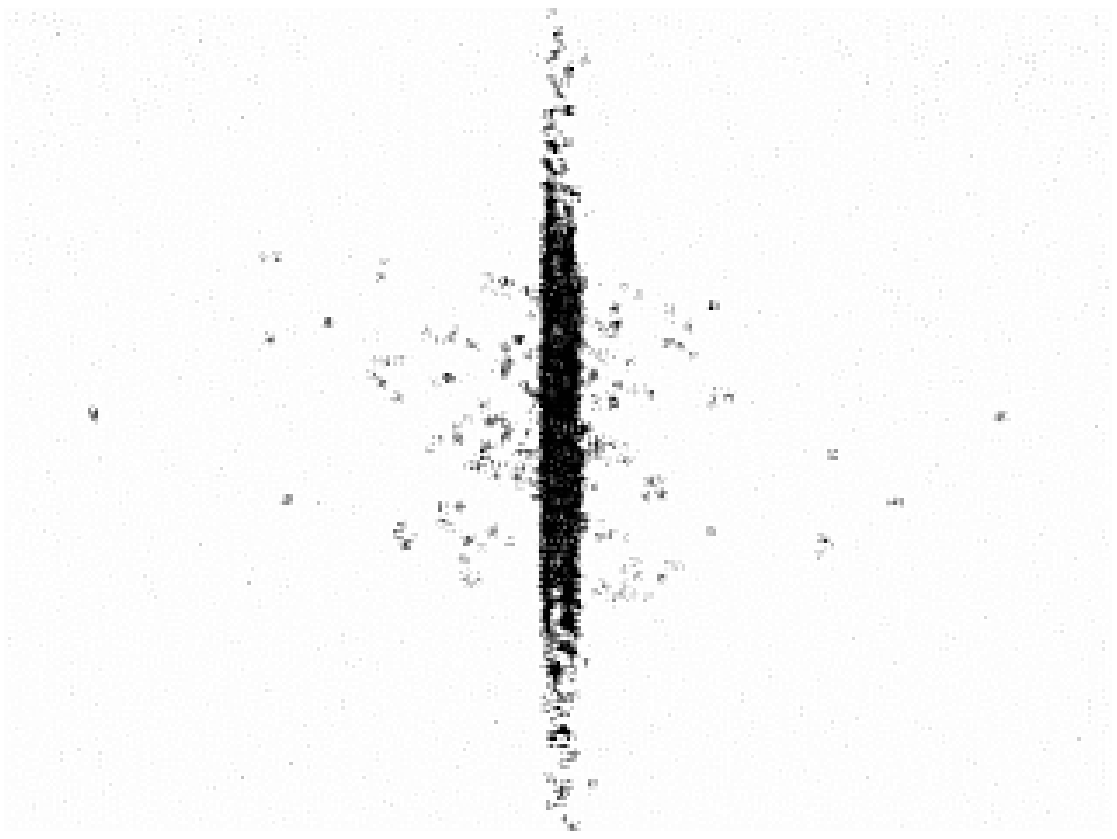}}
\framebox{\includegraphics[width=3.1cm]{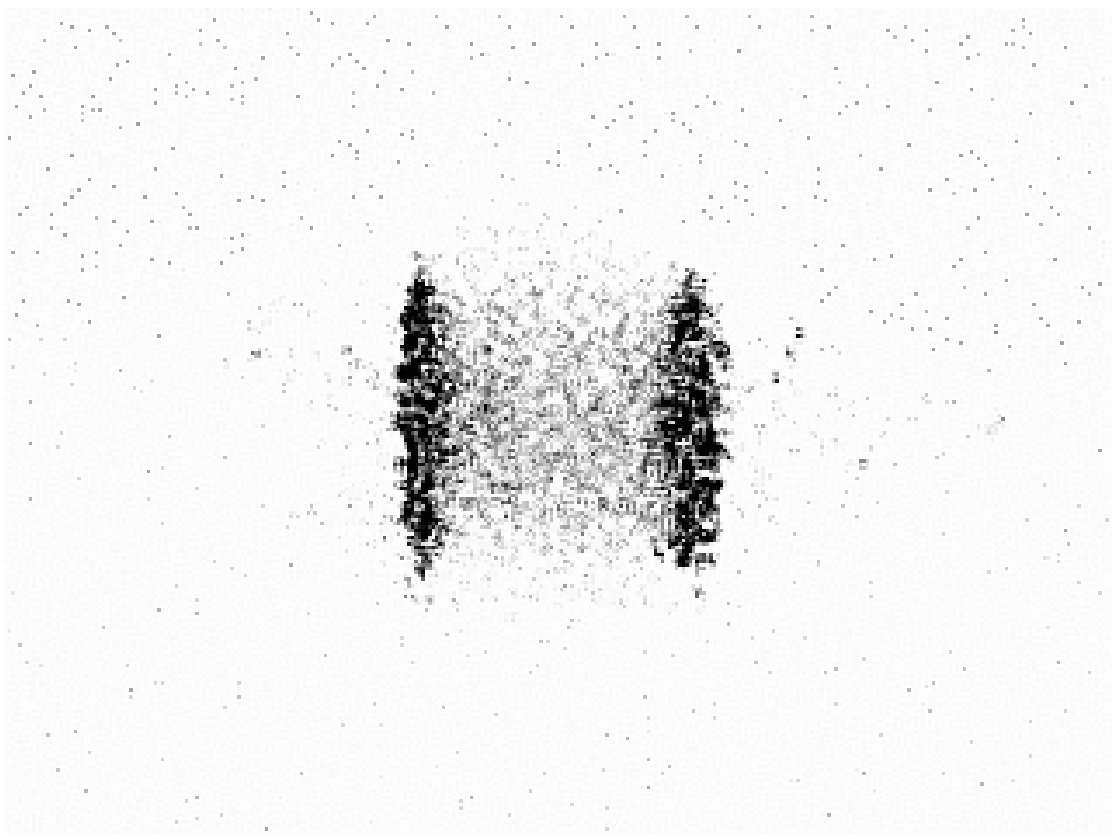}}\\
\framebox{\includegraphics[width=3.1cm]{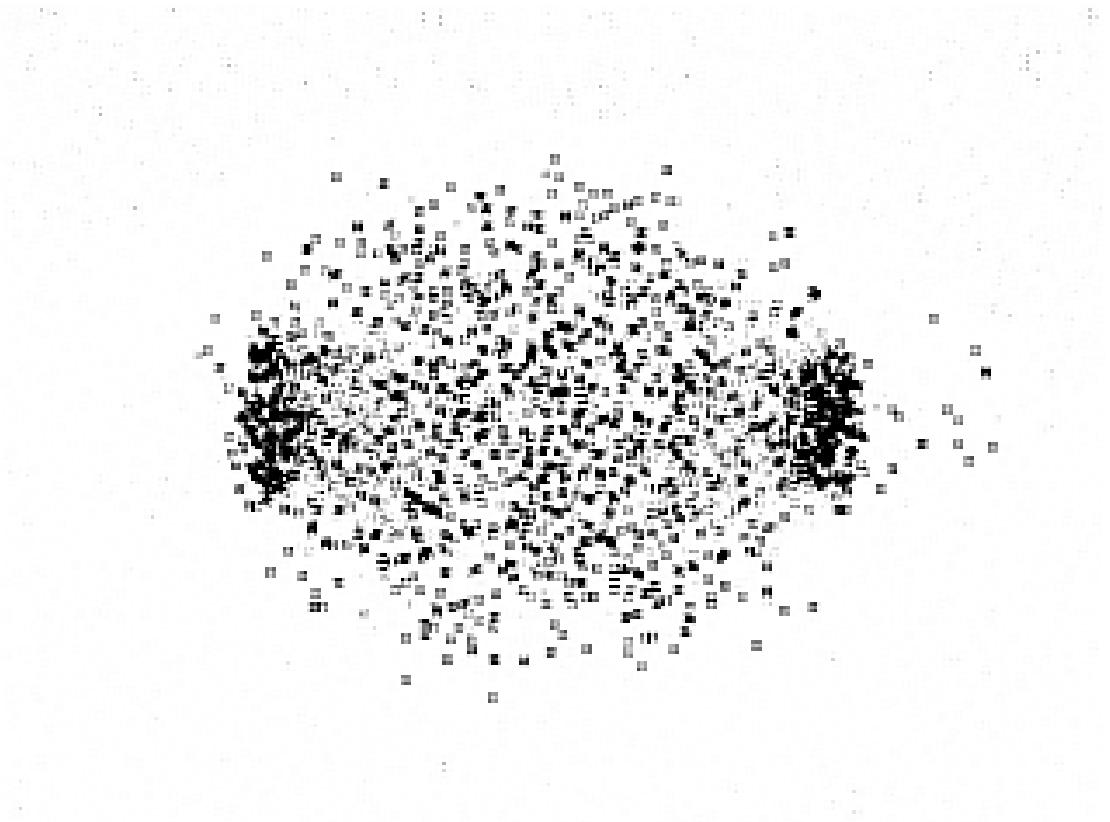}}
\framebox{\includegraphics[width=3.1cm]{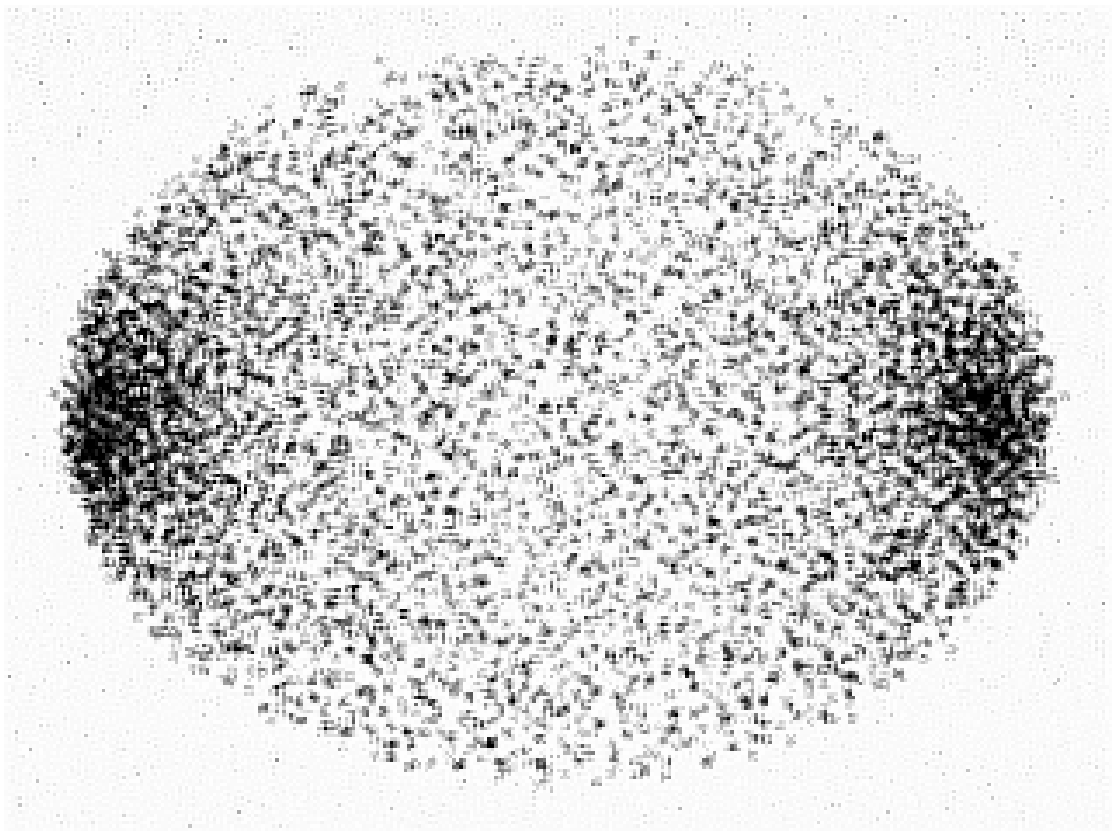}}\\
\end{center}
\vspace*{-1.1cm}
\caption{The four basic stages of an ultra-relativistic
$Au$+$Au$ collision.
For details see Ch.~\ref{ch:1coll}.
}
\end{figure}

\noindent
The baryo-chemical potential $\mu_B$ denotes 
the energy which is necessary to add one nucleon to the system.
Recent lattice QCD calculations 
indicate the position of the tri-critical point
of strongly interacting matter at
a temperature $T$=(160$\pm$3.5)~MeV and a baryo-chemical
potential of $\mu_b$=(725$\pm$35)~MeV \cite{lattice}. 
Thus, the primordial conditions in a RHIC collision
are close to the critical temperature, but 
the baryo-chemical potential has a difference
of a few hundred MeV to the tri-critical point\footnote{
Note that the new GSI accelerator facility FAIR (see Ch.~\ref{ch:dau})
will test a ($\mu_b$,$T$) region in the vicinity of the 
predicted tri-critical point, where event-by-event 
fluctuations might occur.}.
As one consequence, large scale event-by-event fluctuations
in the number of charged hadrons (as one might expect close to 
the tri-critical point) are not expected at RHIC.
However, it should be noted, that in the 2-dimensional
QCD phase diagram ($T$ vs.\ $\mu_b$)
the primordial conditions of a RHIC collision
are very close to those of the primordial universe.

\subsection{The Temperature}

\quad \\
One distinguishes between
the chemical freeze-out temperature $T_{chem}$
and the kinetic freeze-out temperature $T_{kin}$,
which correspond to different times in the system evolution.
$t$($T_{chem}$) denotes the end of the
{\it inelastic} collisions inside the particle wave front,
i.e.\ relative particle abundances become fixed.
$t$($T_{kin}$) denotes the end of the
{\it elastic} collisions and is $\simeq$100~MeV
cooler than $T_{chem}$.
The temperature $T_{chem}$=181~MeV corresponds to 
2.1$\cdot$10$^{12}$~K, which can be compared to
the temperature in other systems (Tab.~3).
The temperature resembles one of the highest
temperatures in the universe. 

\noindent
Also an interesting fact from the comparison 
of the temperatures at $\sqrt{s}$=130~GeV and $\sqrt{s}$=200~GeV
shall be noted: although the collision energy 
increases by $\Delta$$E$=70~GeV, 
the temperature only increases by $\Delta$$T$=7~MeV,
which corresponds to a small fraction of $\leq$$10^{-6}$$A$$\cdot$$\Delta$$E$.

\quad\\
\begin{tabular}{@{}rllll}
\multicolumn{4}{l}{\hspace*{-0.2cm}Table 3}\\
\multicolumn{4}{l}{\hspace*{-0.2cm}Comparison of temperatures in various systems.}\\
\hline
1.4$\cdot$10$^{34}$~K & Planck temperature\\
2.1$\cdot$10$^{12}$~K & $Au$+$Au$ collision\\
$\simeq$$10^{11}$~K & mini black hole ($R$=1~fm, $M$=$10^{12}$~kg)\\
$\simeq$10$^9$~K & supernova\\
15$\cdot$10$^6$~K & sun (core)\\
55$\cdot$10$^6$~K & plasma fusion\\
4$\cdot$10$^6$~K & laser fusion\\
3$\cdot$10$^4$~K & thunderstorm flash \\
\hline
\end{tabular}

\subsection{The Matter Density}

\label{ch:matterdens}

\quad\\
The spatial matter density,
i.e.\ the number of partons per volume
$\rho$=$N_{partons}$/$V$
can be estimated by a Bjorken ansatz \cite{Bjorken}.
If assuming that initial and
final entropy are equal,
the number of partons at $t$$\leq$1~fm/c
is equal to the measured number of final state hadrons $N_{hadron}$.
The volume can be calculated by inserting the fireball radius
($R_{out}$ and $R_{long}$ in Tab.~2)
into a cylindrical volume 
(due to Lorentz boost in the beam direction),
i.e.\ $\rho$$\simeq$$dN_{parton}$/$dy$$\cdot$1/($\pi$$R^2$$t$)
using the rapidity\footnote{The rapidity $y$ 
of a particle is defined as 
$y$=1/2$\cdot$ln(($E$+$p_z$)/($E$+$p_z$)) 
with the total particle energy $E$ and the momentum component
in beam direction $p_z$. The yield $dN$/$dy$ is Lorentz invariant.} $y$.
For a time early in the system evolution 
$t$=0.2~fm/c the matter density is $\rho$$\simeq$20/fm$^3$,
which corresponds to $\simeq$15$\times$$\rho_o$,
$\rho_o$ denoting the density of cold gold nuclei.
At this high density, hadrons are definitely non-existent.
In Tab.~4, the density is compared to other
systems.

\quad\\
\begin{tabular}{@{}rllll}
\multicolumn{4}{l}{\hspace*{-0.2cm}Table 4}\\
\multicolumn{4}{l}{\hspace*{-0.2cm}Comparison of the density in various systems.}\\
\hline
30$\cdot$10$^{17}$~kg/cm$^3$ & $Au$+$Au$ collision \\
2$\cdot$10$^{17}$~kg/cm$^3$ & $Au$ nuclear density \\
$\sim$20,000~kg/cm$^3$ & $Au$ atomic density (solid) \\
$\sim$1000~kg/cm$^3$ & metallic hydrogen \\
 & (core of planet Jupiter)\\
1.1$\cdot$10$^{-26}$~kg/cm$^3$ & universe critical density \\
\hline
\end{tabular}

\subsection{The Energy Density}

\quad\\
About 90\% of all emerging particle are $\pi$ mesons,
thus one may assume (as an educated guess) the $\pi$ meson mass 
in the particle energy $E^2$=$m^2$+$p^2$.
By {\it (a)} counting the number of charged particles,
{\it (b)} measuring the average particle momentum $p$
and {\it (c)} using the volume according to Ch.~\ref{ch:matterdens},
one can estimate the total energy density
as $\varepsilon$$\simeq$5~GeV/$fm^3$,
corresponding to a factor $\simeq$30 higher energy density
then in cold $Au$ nuclei.
For comparison, the estimate 
for the AGS is $\varepsilon$=1.2~GeV/fm$^3$,
for the SPS $\varepsilon$=2.4~GeV/fm$^3$ \cite{andronic}.
A better estimate based upon perturbative QCD can be found
elsewhere \cite{kharzeev2}, leading to an even higher energy 
density of $\varepsilon$$\simeq$18~GeV/fm$^3$.

\vspace*{-0.3cm}

\section{From $p$+$p$ to $Au$+$Au$}

\label{ch:pp2AuAu}

\begin{figure}[hhh]
\vspace*{-0.7cm}
\hspace*{-0.2cm}\includegraphics[angle=270,width=3.5cm]{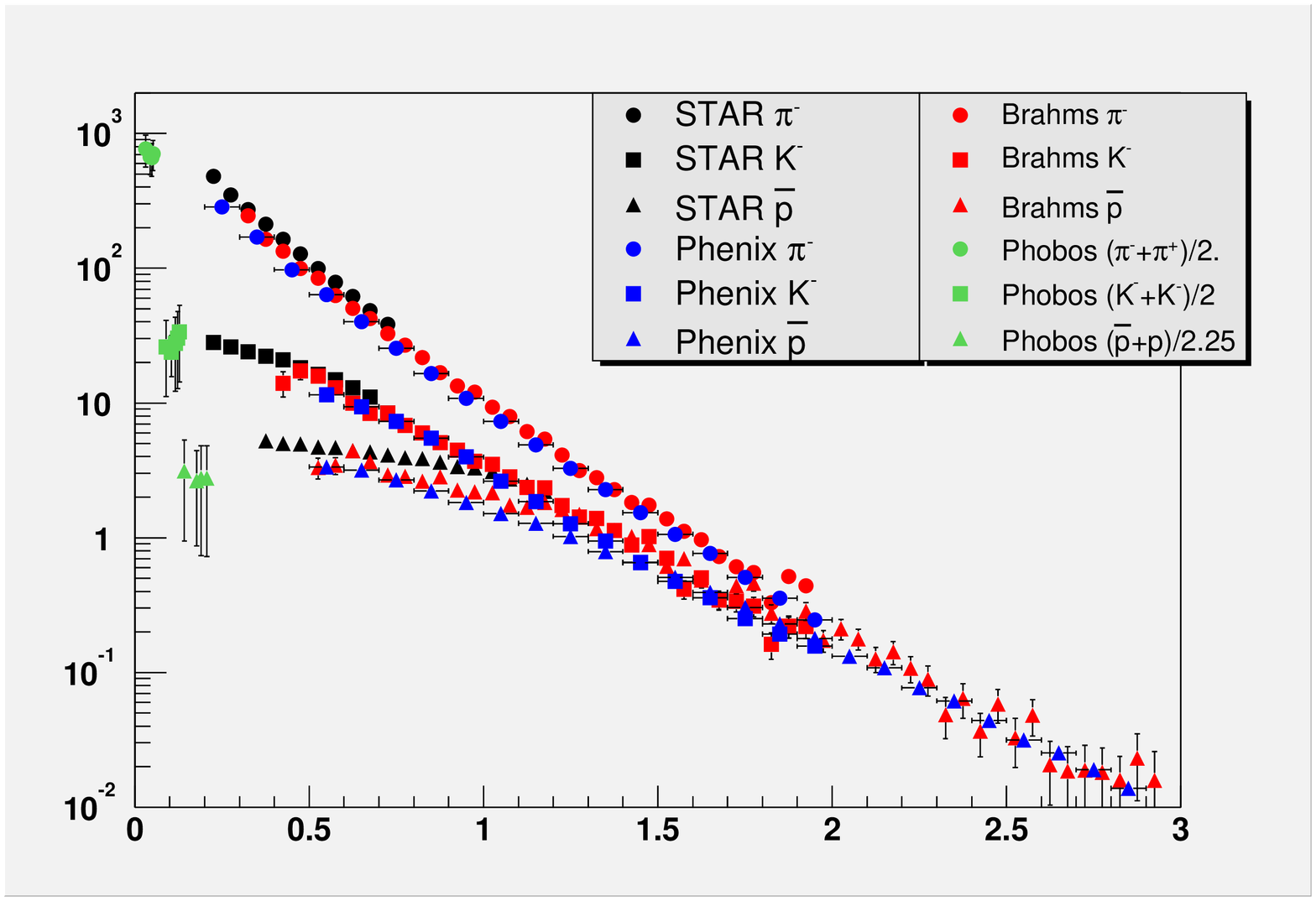}|\\
\hspace*{-0.2cm}\includegraphics[angle=270,width=3.5cm]{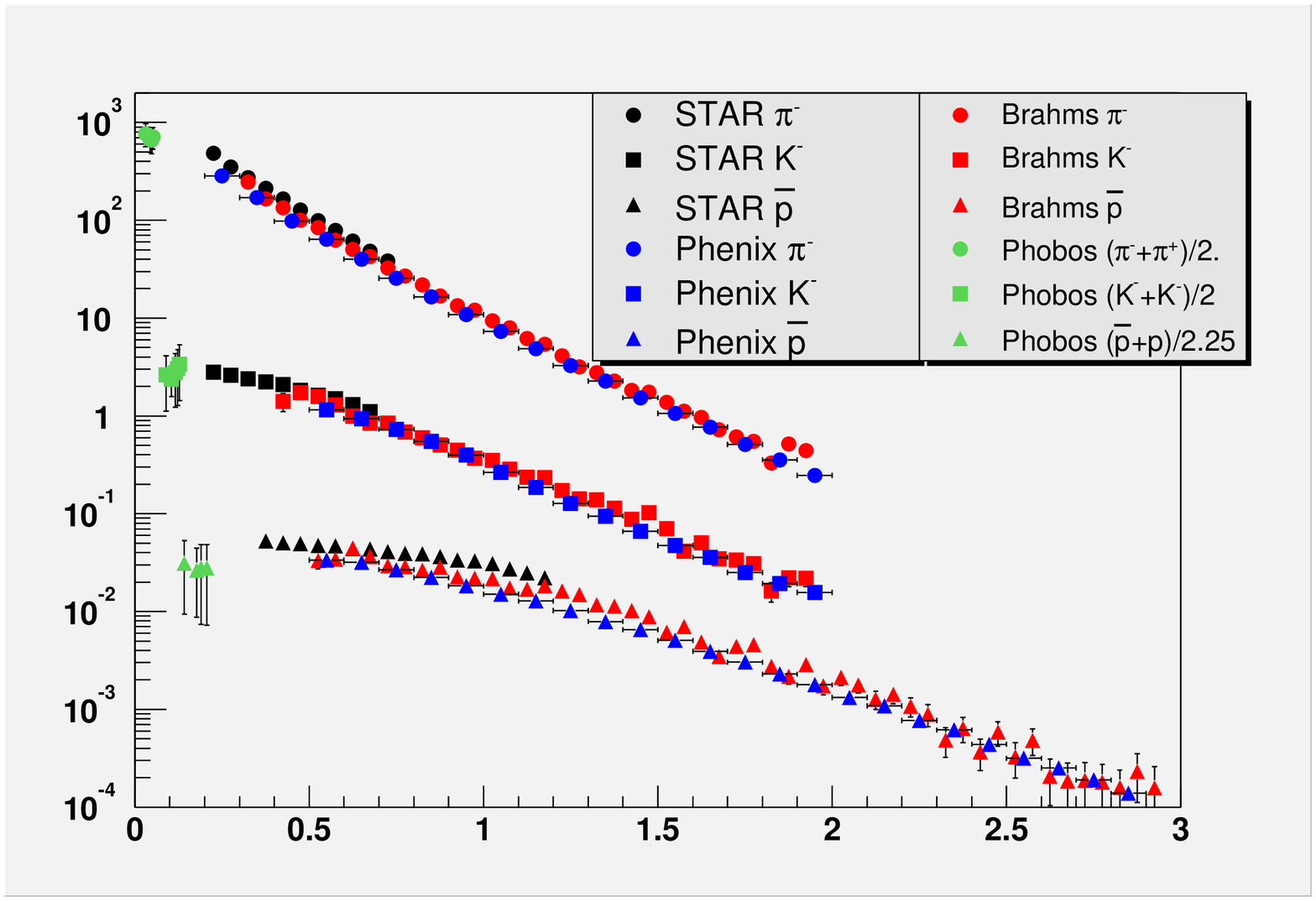}|\\
\vspace*{-0.6cm}
\caption{Transverse momentum distribution 
for $\pi^-$, $K^-$ and $\overline{p}$ 
for $\sqrt{s}$=200~GeV $p$+$p$ (top) and 
$Au$+$Au$ (bottom). For details see Ch.~\ref{ch:pp2AuAu}.
}
\label{f:pp2AuAu}
\end{figure}

\noindent
As RHIC experiments measure both $Au$+$Au$ and $p$+$p$ collisions
in the same detector,
corresponding data of both 
collision types can be directly compared with
identical systematic errors.
Fig.~\ref{f:pp2AuAu} shows transverse momentum distributions
for $\pi^-$, $K^-$ and $\overline{p}$ 
for $\sqrt{s}$=200~GeV $p$+$p$ (top) and 
$Au$+$Au$ (bottom) as a compilation 
of all four RHIC experiments.
One can immediately see that a $Au$+$Au$ collision is not
a simple superposition of many $p$+$p$ collisions.
In fact, the average mean transverse momentum $<$$p_T$$>$ is 
$\simeq$390~MeV/c for $p$+$p$ and
$\simeq$508~MeV/c for central $Au$+$Au$.
The increase by $\simeq$30\% indicates 
the presence of collective effects.
For further details see \cite{pp2AuAu} \cite{ullrich}.

\section{Qualitative comparison to the primordial universe}

\noindent
A RHIC $Au$+$Au$ collision is sometimes being refered to as the ``little bang'',
in reference to the ``big bang'' of the primordial universe.
However, a few important differences shall be noted:

\noindent $\bullet$ 
The primordial universe underwent a matter-dominated 
and a radiation-dominated period.
At RHIC, mat\-ter do\-mi\-nates ex\-clu\-sive\-ly.

\noindent $\bullet$ 
At the time when the universe reached RHIC temperatures of $T$$\simeq$100~MeV, 
it already had developed a macroscopical horizon distance of $L$$\simeq$10~km,
and thus had a size which was roughly larger than the size of a RHIC collision 
by a factor 10$^{18}$.

\noindent $\bullet$ 
The expansion velocity of $v$=0.56$c$ at RHIC 
indicates an almost explosive character.
In case of the primordial universe (after inflation)
it was much slower by a factor
1/$M_{Planck}$$\simeq$10$^{19}$.

\noindent $\bullet$ 
In the system evolution of the universe,
RHIC matter density of $\rho$$\simeq$15$\rho_o$ 
occured only for a very short period
{\it during} inflation (assuming inflation 
from $10^{+118}$$\rho_o$ at $t$$\sim$$10^{-36}$~s 
to $10^{-19}$$\rho_o$ at $t$$\sim$$10^{-34}$~s).
Additional details are given elsewhere \cite{soeren_primordial}.

\section{Anti-Helium}

\label{ch:antihe}

\noindent
Currently one of the most interesting questions of cosmology 
is: {\it where is the primordial anti-matter}~?
Several experiments (e.g.\ 
the balloon bound BESS experiment \cite{bess} at a height of $H$=27~km 
and the space-shuttle bound AMS experiment \cite{ams} at an orbit
of $H$=320-390~km) search for direct evidence of anti-matter 
in cosmic rays.
It should be noted that incident particle energies 
at the AMS orbit location (i.e.\ quasi-``beam'' energies) range 
from 10$^1$-10$^{6}$~MeV per nucleon, comparable to
RHIC beam energies at 10$^{5}$~MeV per nucleon.
A par\-ti\-cu\-lar\-ly suited probe for anti-matter search
is anti-$He$, as the signature of charge $Z$=$-$2
is clean\footnote{Except anti-$He$ and hypothetical pentaquark
states there are no other stable $Z$=$-$2 particles predicted in nature.}.
A $Z$=$-$1 signature of $\overline{p}$ is contaminated by 
$K^-$ and $\pi^-$ mesons. 
The STAR level-3 trigger system was used to identify
$Z$=$-$2 candidate events in realtime.
Primordial Helium is not rare, but in fact represents
$\simeq$24\% of all primordially created matter
(created at time scales $t$$\sim$1~s and at
temperatures $T$$\sim$0.1~MeV). 
Under the assumption of an {\it a priori} symmetric primordial production 
of $He$ and anti-$He$, one may conclude that anti-$He$ 
might be detectable, too.

\noindent
In a RHIC $Au$+$Au$ collision,
there is trivially no anti-matter in the initial state,
as nuclei are completely made of matter.
However, even anti-nuclei are created in a detectable quantity.
The dominant anti-nuclei production mechanism is a two-step process, namely
{\it (1)} pair production of $p$$\overline{p}$, $n$$\overline{n}$,
followed by {\it (2)} coalescence, i.e.\
the anti-nucleon wave functions overlap inside a homogenity volume.
For the systematic study of the $A$ dependence of anti-nuclei 
production yields,
it is useful to define an {\it invariant yield}

\vspace*{0.2cm}
\begin{equation}
E \frac{d^3 N_A}{d^3 P} = B_A
( E \frac{d^3 N_N}{d^3 p} )^A
\end{equation}
\vspace*{0.2cm}

\noindent
with anti-nuclei yield $N_A$, 
the anti-nucleon yield $N_N$,
the anti-nucleon momentum $p$, 
and $p$=$P$/$A$.
The coalescence coefficient $B_A$ denotes the probability that $A$ anti-nucleons form a bound state.
It might be regarded as a ``penalty factor'' for the step from $n$ to $n+1$ anti-nucleons
in the system. Typical orders of magnitude are $B_2$$\sim$10$^{-3}$ and $B_3$$\sim$10$^{-6}$.
For high $\sqrt{s}$ and a large system size, the coalescence coefficient is related
to the inverse of the effective volume containing the anti-nuclei (i.e.\ the fireball
volume) by $B_A$=1/$V_{eff}^{A-1}$.

\noindent
So far, STAR has been able to collect a raw yield of 193 anti-$^3He$
in a total of 4$\cdot$10$^6$ $Au$+$Au$ events at $\sqrt{s}$=200~GeV.
In the same data set, 6416 anti-deuterons were also identified
and used for a determination of $B_2$.
Further details are given elsewhere \cite{antihe_star}.
Fig.~5 shows the measured ionization $dE/dx$ 
vs.\ momentum for the anti-$He$ candidates 
in the STAR Time Projection Chamber.

\begin{figure}[hhh]
\label{f:antihe}
\centerline{\includegraphics[width=80mm,height=55mm]{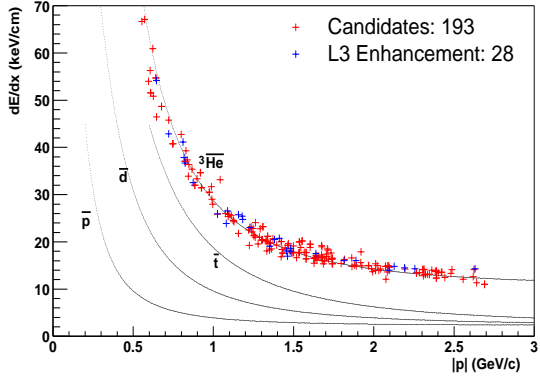}}
\vspace*{-0.8cm}
\caption{Ionization $dE/dx$ vs.\ momentum $p$ for anti-$He$
in the STAR Time Projection Chamber ($blue:$ level-3 trigger, $red:$ offline analysis).}
\end{figure}

\noindent
Fig.~6 shows the extracted coalescence coefficient $B_3$
in comparison to other measurements. RHIC represents the highest
energy ever at which anti-$He$ was created in a accelerator experiment. 

\begin{figure}[hhh]
\label{f:b3}
\centerline{\includegraphics[width=80mm,height=55mm]{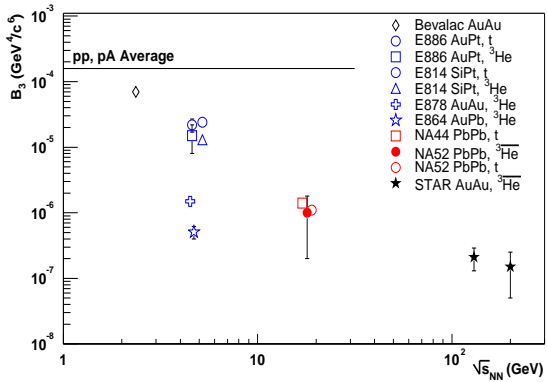}}
\vspace*{-0.8cm}
\caption{Coalescence coefficient $B_3$ for anti-$^3He$ and anti-$t$
production vs.\ $\sqrt{s}$ for different experiments.}
\end{figure}

\noindent
As an important result, the coalescence (and thus the yield of produced
anti-nuclei per one collision) in $Au$+$Au$ (normalized to the number
of binary collisions) at RHIC is approximately a factor of 40
smaller then in $p$+$p$ or in $\gamma$+$p$ \cite{h1}.

\noindent
The coalescence coefficient is also an important input to calculations
of anti-$He$ yield from nucleus-nucleus collisions in the 
stellar medium \cite{chardonnet}, which is the dominant background 
for the primordial anti-$He$ search by AMS.

\noindent
The STAR 2004 data set is currently being analyzed.
The data are expected to contain 2.7$\pm$0.3 anti-$^4$$He$, 
which would be a first experimental observation.

\section{Forward Physics}

\label{ch:dau}

\noindent
For high energy $p$+$A$ collisions in the earth's upper atmosphere,
a proper knowledge of the particle shower in the forward rapidity region 
is very important, as the forward particles are the visible center of
the showers as seen on the detector systems at sea level.
Ideally $p$+$Au$ collisions would be preferable, but due to technical 
accelerator issues\footnote{One the one hand, the $d$ beam has a charge/mass 
ratio close to the $Au$ beam. On the other hand the beam-beam interaction, which shortens
the lifetime of a RHIC store, is in the case of $p$+$Au$ a factor of $\simeq$2
higher than in the case of $d$+$Au$.} $d$+$Au$ collisions were chosen at RHIC.
As shown in Tab.~1, several million $d$+$Au$ collisions were 
recorded at RHIC in 2003. 
Fig.~7 shows the rapidity distribution of particles from
a $d$+$Au$ collision at $\sqrt{s}$=200~GeV with data from BRAHMS \cite{dau_brahms}, 
in comparison to a perturbative QCD calculation \cite{kharzeev1}.
The $\eta$ asymmetry ratio of forward ($Au$ side) to backward ($d$ side)
is a characteristic of the collision system.
STAR measured the ratio for these collisions
as a function of $p_T$ (Fig.~8) \cite{dau_star}.
As a surprising result, for 0$\leq$$|$$\eta$$|$$\leq$0.5 the ratio is flat,
but for 0.5$\leq$$|$$\eta$$|$$\leq$1.0 the ration 
shows a rising behaviour up to $p_T$$\simeq$2-3~GeV/c,
followed by a descending slope for higher $p_T$. Fig.~8 also
shows the comparison to model calculations, including 
{\it (a)} shadowing\footnote{Shadowing is the depletion of parton distributions at small $x$ 
inside a nucleus.} and {\it (b)} saturation at high gluon densities.
The data seem to support a saturation ansatz, which, in a geometrical picture, 
assumes that the probability that two gluons collide is one.
$p$+$A$ collisions will also be an important part of the future GSI accelerator 
facility FAIR at Darmstadt, Germany (Fig.~9), which will provide $p$ beams
for fixed target experiments up to 90~GeV.

\begin{figure}[hhh]
\label{f:dau1}
\vspace*{0.3cm}
\rightline{\includegraphics[height=4.6cm,width=6.5cm]{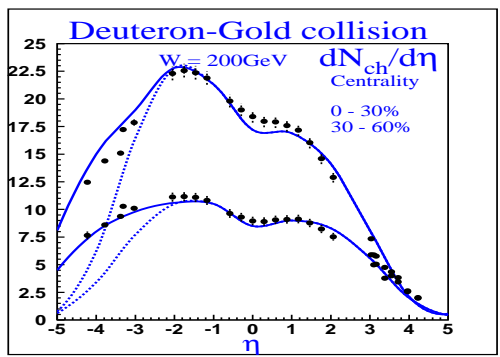}}
\vspace*{-0.7cm}
\caption{Pseudorapidity distribution for charged particles from
$d$+$Au$ collisions at $\sqrt{s}$=200~GeV in BRAHMS \cite{dau_brahms}
for two different centralities, corresponding to two different impact parameter ranges.}
\end{figure}

\begin{figure}[hhh]
\label{f:dau2}
\vspace*{0.2cm}
\rightline{\hspace*{0.5cm}\includegraphics[width=7.2cm]{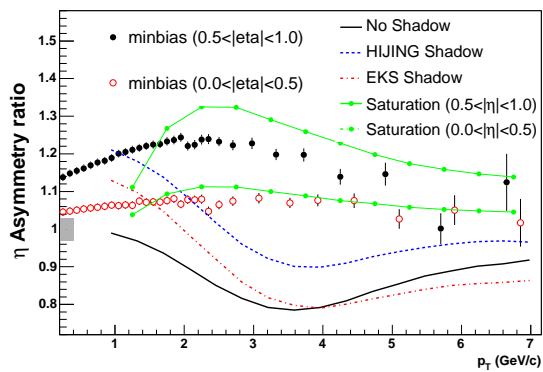}}
\vspace*{-0.8cm}
\caption{Forward/backward pseudorapidity ratio vs.\ $p_T$
$d$+$Au$ collision at $\sqrt{s}$=200~GeV in STAR \cite{dau_star}.
For details see Ch.~\ref{ch:dau}.}
\end{figure}

\begin{figure}[hhh]
\label{f:fair}
\vspace*{0.1cm}
\centerline{\includegraphics[width=5.5cm,angle=270]{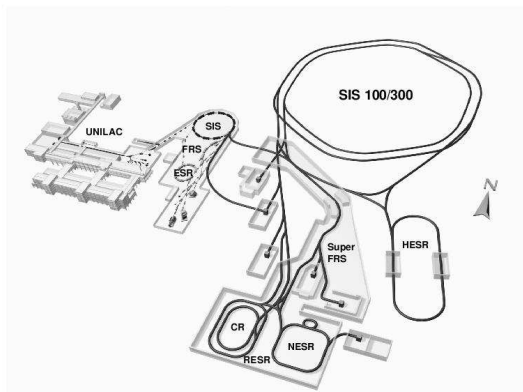}}
\vspace*{-0.7cm}
\caption{The future GSI accelerator complex FAIR.}
\end{figure}

\begin{figure}[hhh]
\label{f:rho}
\vspace*{0.2cm}
\centerline{\includegraphics[height=5cm,width=7.7cm]{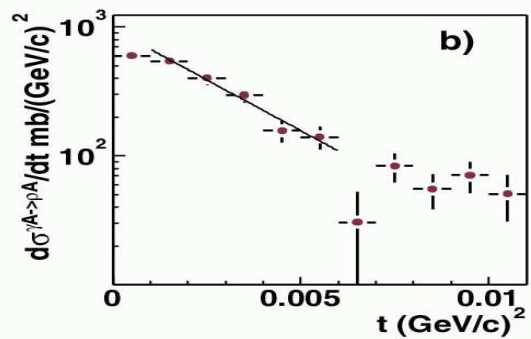}}
\vspace*{-0.9cm}
\caption{Forward differential cross section 
$d$$\sigma$($\gamma$$Au$$\rightarrow$$\rho^o$$Au$)/$dt$ vs.\ t \cite{star_rho}.
}
\end{figure}

\noindent
Another important class of forward reactions are ultra-peripheral $Au$+$Au$
collisions, in which the $Au$ nuclei interact by a very strong 
electromagnetic field (10$^{20}$~V/cm at the surface of the nuclei).
Photons produced in these strong fields are peaked forward,
and may create vector mesons by scattering off the $Au$ nuclei.
Fig.~10 shows the differential cross section 
$d$$\sigma$($\gamma$$Au$$\rightarrow$$\rho^o$$Au$)/$dt$ vs.\ the momentum transfer $t$.
The very small $t$ values indicate the forward scattering. 
From an exponential fit (shown as line in Fig.~10) 
a forward cross section was determined to be
$d$$\sigma$/$dt$$|_{t=0}$=965$\pm$140$\pm$230~mb/GeV$^2$.
Details are given elsewhere \cite{star_rho}.

\section{Charm cross sections}

\noindent
For high energy cosmic nuclear collisions in the galactic medium or 
the earth's atmosphere, the knowledge of heavy quark production
cross sections is very important.
Both STAR \cite{charm_star} and PHENIX \cite{charm_phenix} have measured
the charm cross section.
Fig.~11 shows the total $c$$\overline{c}$ cross section 
per nucleon-nucleon collision vs.\ $\sqrt{s}$. 
There is a preliminary indication that the measured cross section
is higher than the expected cross section extrapolation 
obtained from 
{\it (a)}~PYTHIA Monte-Carlo simulation
(based upon Lund fragmentation) \cite{pythia} and 
{\it (b)}~a NLO QCD calculation \cite{charm_NLO}.

\begin{figure}[hhh]
\label{f:charm}
\centerline{\includegraphics[width=7.5cm]{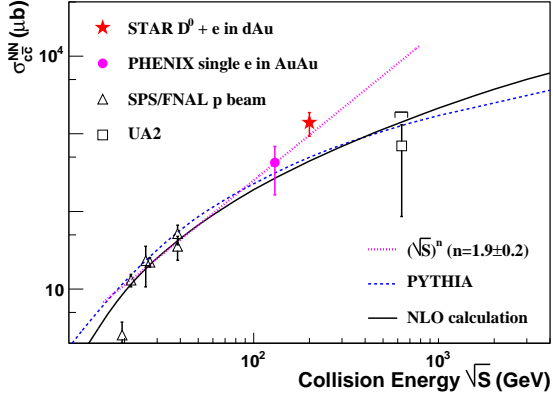}}
\vspace*{-0.6cm}
\caption{Total $c$$\overline{c}$ cross section 
per nucleon-nucleon collision vs.\ $\sqrt{s}$. For details see \cite{charm_star}.}
\label{fig:dau1}
\end{figure}

\section{Summary}

\noindent
In 4 years of operation, RHIC has collected a data set of millions
of $p$+$p$, $d$+$Au$ and $Au$+$Au$ collisions. From the analysis we may
conclude that a RHIC $Au$+$Au$ collision is a very hot and very dense 
hadronic system, which expands explosively with $v$=0.56$c$.
During the short freeze-out the system undergoes a graceful QCD self-organisation,
which could be successfully described by models based upon perturbative QCD. 
One of the most basic but also most important observations is
that a $Au$+$Au$ collisions is not a
simple superposition of $p$+$p$ collisions, but in fact the mean transverse
momentum increases by $\simeq$30\%, indicating collective effects.


\begin{thebibliography}{9}

\bibitem{rhic_overview}
{\tt http://www.bnl.gov/rhic}

\bibitem{star}
{\tt http://www.star.bnl.gov}




\bibitem{phenix}
{\tt http://www.phenix.bnl.gov}

\bibitem{phobos}
{\tt http://www.phobos.bnl.gov}

\bibitem{brahms}
{\tt http://www.rhic.bnl.gov/brahms}

\bibitem{star_l3}
J.~S.~Lange et al.,\\
Nucl.~Instr.~Meth.~{\bf A499}(2003)778

\bibitem{kharzeev2}
D.~Kharzeev, M.~Nardi, 
nucl-th/0012025,\\
Phys.~Lett.~{\bf B507}(2001)121

\bibitem{star_Et}
The STAR Collaboration, 
nucl-ex/0407003

\bibitem{star_hbt}
The STAR Collaboration, 
nucl-ex/0107008,\\
Phys.~Rev.~Lett.~{\bf 87}(2001)0C82301

\bibitem{star_multistrange}
The STAR Collaboration, 
nucl-ex/0307024,\\
Phys.~Rev.~Lett.~92(2004)182301

\bibitem{star_lee2}
The STAR Collaboration, 
nucl-ex/0310004,\\
Phys.~Rev.~Lett.~92(2004)112301,

\bibitem{star_mub}
The STAR Collaboration, 
nucl-ex/0211024,\\
Phys.~Lett.~{\bf B567}(2003)167,

\bibitem{star_antip}
The STAR Collaboration, \mbox{nucl-ex/0104022},\\
Phys.~Rev.~Lett.~{\bf 86}(2001)4778,\\
Err.~Phys.~Rev.~Lett.~{\bf 90}(2003)119903(E)

\bibitem{star_lee1}
The STAR Collaboration, 
nucl-ex/0311017

\bibitem{centauro}
M.~Tamada et al., Nuovo~Cim. B41(1977)245

\bibitem{lattice}
Z.~Fodor, S.~D.~Katz,
hep-lat/0106002,\\
JHEP 0203(2002)014

\bibitem{Bjorken}
J.~D.~Bjorken, 
Phys.~Rev.~{\bf D27}(1983)140

\bibitem{andronic}
A.~Andronic, P.~Braun-Munzinger,\\
hep-ph/0402291

\bibitem{pp2AuAu}
The STAR Collaboration,
nucl-ex/0310004,\\
Phys.~Rev.~Lett.~{\bf 92}(2004)112301

\bibitem{ullrich}
The STAR Collaboration,
nucl-ex/0211004,\\
Nucl.~Phys.~{\bf A715}(2003)399

\bibitem{soeren_primordial}
J.~S.~Lange, hep-ph/0403104 

\bibitem{bess}
The BESS Collaboration,
astro-ph/9710228\\
Phys.~Lett.~{\bf B422}(1998)319

\bibitem{ams}
The AMS Collaboration, 
hep-ex/0002048,\\
Phys.~Lett.~{\bf B472}(2000)215,

\bibitem{antihe_star}
J.~S.~Lange, C.~Struck,
nucl-ex/0403008,\\
Nucl.~Phys.~{\bf A738}(2004)396

\bibitem{h1}
The H1 Collaboration,
hep-ex/0403056,\\
Eur.~Phys.~J.~{\bf C36}(2004)413

\bibitem{chardonnet}
P.~Chardonnet, J.~Orloff, P.~Salati,\\
astro-ph/9705110,\\
Phys.~Lett.~{\bf B409}(1997)313,

\bibitem{dau_brahms}
The BRAHMS Collaboration,\\
nucl-ex/0401025

\bibitem{kharzeev1}
D. Kharzeev, E. Levin, M. Nardi,\\
hep-ph/0212316,
Nucl.~Phys.~{\bf A730}(2004)448

\bibitem{dau_star}
The STAR Collaboration, 
nucl-ex/0408016

\bibitem{star_rho}
The STAR Colaboration,
nucl-ex/0206004,
Phys.~Rev.~Lett.~89(2002)272302

\bibitem{charm_star}
The STAR Collaboration,
nucl-ex/0407006

\bibitem{charm_phenix}
The PHENIX collaboration,\\
nucl-ex/0202002, Phys.~Rev.~Lett.~{\bf 88}(2002)192303

\bibitem{pythia}
T.~Sj\"ostrand et al.,\\
Comp.~Phys.~Comm.~{\bf 135}(2001)238 

\bibitem{charm_NLO} 
R.~Vogt, 
hep-ph/0203151,\\
Int.~J.~Mod.~Phys.~{\bf E12}(2003)211





\end{thebibliography}
\end{document}